\documentclass[twocolumn,amsmath,amssymb,aps,english,superscriptaddress,nofootinbib]{revtex4}
\usepackage{graphicx,amssymb,amsmath}
\usepackage{babel}

\begin{document}

\title{Market impact and trading profile of large trading orders in stock markets}
\author{Esteban Moro}
\affiliation{Instituto de Ciencias Matem\'aticas CSIC-UAM-UC3M-UCM, Madrid, Spain}
\affiliation{Grupo Interdisciplinar de Sistemas Complejos (GISC)
and Departamento de Matem\'aticas, Universidad Carlos III de
Madrid, E-28911, Legan\'es (Madrid), Spain}

\author{Javier Vicente}
\affiliation{Santa Fe Institute, 1399 Hyde Park Road, Santa Fe, NM 87501}

\author{Luis G. Moyano}
\affiliation{Grupo Interdisciplinar de Sistemas Complejos (GISC)
and Departamento de Matem\'aticas, Universidad Carlos III de
Madrid, E-28911, Legan\'es (Madrid), Spain}

\author{Austin Gerig}
\affiliation{Santa Fe Institute, 1399 Hyde Park Road, Santa Fe, NM 87501}

\author{J. Doyne Farmer}
\affiliation{Santa Fe Institute, 1399 Hyde Park Road, Santa Fe, NM 87501}
\affiliation{LUISS Guido Carli, Viale Pola 12, 00198, Roma, Italy}

\author{Gabriella Vaglica}
\affiliation{Dipartimento di Fisica e Tecnologie Relative, Universit\`a di Palermo, 
Viale delle Scienze, I-90128 Palermo, Italy}

\author{Fabrizio Lillo}
\affiliation{Santa Fe Institute, 1399 Hyde Park Road, Santa Fe, NM 87501}
\affiliation{Dipartimento di Fisica e Tecnologie Relative, Universit\`a di Palermo, 
Viale delle Scienze, I-90128 Palermo, Italy}
\affiliation{CNR-INFM-SOFT, Roma, Italy}

\author{Rosario N.\ Mantegna}
\affiliation{Dipartimento di Fisica e Tecnologie Relative, Universit\`a di Palermo, 
Viale delle Scienze, I-90128 Palermo, Italy}
\affiliation{CNR-INFM-SOFT, Roma, Italy}

\begin{abstract}
We empirically study the market impact of trading orders.  We are specifically interested in large trading orders that are executed incrementally, which we call {\it hidden orders}.  These are reconstructed  based on information about market member codes using data from the Spanish Stock Market and the London Stock Exchange.  We find that market impact is strongly concave, approximately increasing as the square root of order size.  Furthermore, as a given order is executed, the impact grows in time according to a power-law; after the order is finished,  it reverts to a level  of about $0.5-0.7$ of its value at its peak. We observe that hidden orders are executed at a rate that more or less matches trading in the overall market, except for small deviations at the beginning and end of the order.
\end{abstract}
\maketitle

\section{Introduction}

Market impact is the expected price change conditioned on initiating a trade of a given size and a given sign.  One naturally expects that initiating a buy order should drive the price up, and initiating a sell order should drive it down.  This has roots in standard economic theory:  An increase in demand should increase prices, while an increase in supply should decrease prices.  Market impact is important for theoretical and practical reasons.   First of all, in order to be able to estimate transaction costs, and in order to optimize a trading strategy to minimize such costs, it is necessary to understand the functional form of market impact \cite{kissel}. Moreover since impact is a cost of trading, it exerts selection pressure against a fund becoming too large, and therefore is potentially important in determining the size distribution of funds \cite{berk,schwarzkopf}. Finally, market impact reflects the shape of excess demand, which is of central importance in economics.
Despite its conceptual and practical importance, a proper empirical characterization and theoretical understanding of market impact is still lacking \cite{bfl}.   

The functional forms of market impact vary from study to study.  This is in part because there are several different types of market impact that must be distinguished.  Studies of individual transaction impact yield a strongly concave functional form  which appears to vary from market to market \cite{lillo,potters,farmer,farmerpnas}. Other studies have looked at market impact under aggregation, in which the impact is conditioned on the sum of the signed transaction volume associated with a given number of trades or a given interval of time.   These studies have tended to observe a somewhat less strongly concave shape \cite{bfl,kempf,plerou,gabaix,hopman,aggregate}.  Other research has focused on orders executed through specific mechanisms, e.g. block markets \cite{keim}. 

In this paper we instead focus on the impact of large trading orders that are split into pieces and executed incrementally, which we call {\it hidden orders}.   For strategic reasons traders attempt to keep the true size of their orders secret in order to minimize transaction cost.  Consider for example a trader who wishes to buy a large number of shares of a company in order to profit on an expected future price increase.  She of course wishes to buy her shares at the lowest price possible.  As she buys she will push the price up.  However, by executing her order incrementally she should be able to buy at least part of the order at a low price, before the impact has been fully felt, thereby lowering her overall cost.  The strategic reasons for this behavior have been studied in a model by Kyle \cite{kyle}, whose theory predicts a linear dependence of impact on trading volume.  More recent theoretical approaches \cite{bertsimas,almgren,almgren2,gabaix,bouchaud,huberman,obizhaeva,gerig,bfl,henri} have proposed different functional forms for the market impact of hidden orders.

Here we study the market impact of hidden orders as a function of trading size and other properties, such as the style of execution.  We also study the temporal dynamics of the impact as the order develops and after it has been completed, and characterize statistical properties such as the time profile for execution.  There are so far only a few studies of the market impact of hidden orders \cite{barra,lakonishok,almgrenrisk}. The main reason for this is that to study hidden orders one must have the identity of the trading accounts originating the orders, and data sets that contain this information are difficult to obtain. The few empirical studies in the literature considered a small set of hidden orders executed by a specific financial institution who made the data available to the authors of the paper. Such studies suffer from the problem that they consider only a fraction of the total number of hidden orders, which may be idiosyncratic in character.  This can also result in small sample sizes and therefore large statistical fluctuations.

In this paper we follow a different approach and attempt to reconstruct all large hidden orders using a statistical method that makes use of information about brokerage firms,  recently developed in reference \cite{vaglica}.  For our data sets each transaction is labeled with codes identifying the members of the exchange who made the transaction.  In most cases the member is acting as a broker, i.e. they are handling a trade for another institution who is not a member of the exchange.  In other cases the member may trade for their own account.  Thus a single  membership code may lump together trades from many different institutions, originated by many different trading strategies.  As a consequence several hidden orders from different trading accounts may be active with the same broker at the same time, making it impossible to be certain that all orders are correctly identified. 

The detection problem is aided by the fact that the size of hidden orders has a heavy-tailed distribution, and large hidden orders can cause dramatic changes in the rate at which a given membership code participates in trades to either buy or to sell.   The method we use was originally developed  for detection of local stationary regions in physiological time series  \cite{bernaola}.  As we use it here, it looks for time intervals in the time series of orders coming through a given membership code when the firm acts as a net buyer or seller at an approximately constant rate. As in reference \cite{vaglica} we interpret these series of trades as belonging to hidden orders.  This method can never be perfectly accurate, but a variety of tests performed in reference \cite{vaglica} suggest that the reconstruction is good enough to recover the most important statistical properties of hidden orders reasonably well.  An important advantage of this approach is that we are able to study large samples of hidden orders coming from the whole market rather than only a subset belonging to a specific institution. 

The paper is organized as follows. In Section \ref{sec:data} we discuss our data sets, the algorithm for detecting hidden orders and the investigated variables. In Section \ref{sec:statistics} we discuss the statistical properties of the variables characterizing the hidden orders. In Section \ref{sec:impact} and \ref{sec:profile} we present our empirical results on the impact of hidden orders and on their trading profile, respectively. Section \ref{sec:conclusions} concludes.

\section{Data and investigated variables}\label{sec:data}

Our databases contain the on-book (SETS) market transactions of the London Stock Exchange (LSE) from January 2002 to December 2004 and the electronic open-book market (SIBE) of the Spanish Stock Exchange (BME, Bolsas y Mercados Espa\~noles) from January 2001 to December 2004. Roughly  $62\%$ of the transactions at the LSE are executed in the open book market and roughly $90\%$ of the transactions at the BME are executed in the electronic market.

We have initially considered a subset consisting of the most heavily traded stocks in the two markets,  $74$ stocks traded in the LSE and $23$ stocks traded in the BME. For both markets we have considered exchange members who made at least one trade per day for at least $200$ trading days per year and with a minimum of $1000$ transactions per year. This filter yielded approximately $60$ exchange member firms per stock.  We then applied the algorithm for detecting hidden orders described in Ref. \cite{vaglica}, which we have already discussed, to identify hidden orders that consist of at least ten transactions. It is worth noting that the detected patches are not necessarily composed of the same type of trades (buy or sell) but that at least $75\%$ of the transacted volume in the patch must have the same sign.
The algorithm detected 90,393 hidden orders in the LSE and 55,309 in the BME.

This study is based entirely on trades that take place through a continuous double auction.  ``Continuous" refers to the fact that trading takes places continuously and asynchronously, and ``double" to the fact that both buyers and sellers are allowed to place and cancel orders at any time.  There are two fundamentally different ways to execute an order in such a market.  One is to use a limit order, in which an order is placed inside the order book, which is essentially a list of unexecuted orders at different prices.  The other is to place a market order, which we define as any order that results in an immediate transaction.  Every transaction involves a market order transacting against a limit order.  A given real order might act as both, e.g. part of it might result in an immediate transaction and part of it might be left in the order book.  We only consider transactions, so in the example above we would treat the first part as a market order and treat the second part as a limit order, but the second part will enter our analysis only if it eventually results in a transaction. The LSE database allows us to identify whether the initiator of the transaction was the buyer or the seller. For BME this information is not available and we infer it with the Lee and Ready algorithm \cite{leeready}

A hidden order is characterized by several variables. These are 
\begin{itemize}
 \item{The execution time  $T$ (in seconds) of the hidden order, measured as the trading time interval between the first and the last transaction of the hidden order.} 
 \item{The number $N$ of transactions of the hidden order. We consider hidden orders of length $N>10$.} 
 \item{The volume $V$ of the hidden order defined as 
\begin{equation}
V = \sum_{j=1}^N v_j,
\end{equation}
where  $v_j$ is the signed volume of each transaction of the hidden order. For buy trades $v_i > 0$ and for sell trades $v_i < 0$. We consider the hidden order to be a buy order if $V>0$ and a sell order if $V<0$. The buying/selling nature of a hidden order is thus encoded in its sign, $\epsilon = \mathrm{sign} (V)$. The volume is the product of the number of shares times the price and is measured in Pounds (LSE) or in Euro (BME).}  
\item{The volume fraction of market orders $f_{mo}$. A hidden order can be implemented with very different liquidity strategies, i.e. with different compositions of market and limit orders. In order to quantify this we define the fraction (in volume) of market orders within a hidden order as
\begin{equation}
f_{mo} = \frac{\sum_{j=1}^N |v_{j,mo}|}{\sum_{j=1}^N |v_j|},
\end{equation}
where $v_{j,mo}$ is the traded volume at each transaction done through market orders. Values of $f_{mo}$ close to zero mean that the broker completed the hidden order by using mainly limit orders, while values of $f_{mo}$ close to one imply the broker used mainly market orders during the execution of the hidden order. } 
\item{The participation rate $\alpha$ of a hidden order defined as
\begin{equation}
\alpha = \frac{\sum_{i=1}^N |v_i|}{V_M},
\end{equation}
where $V_M$ is the unsigned volume of the stock traded in the market  concurrently with the hidden order. Values of $\alpha$ close to zero imply the hidden order was negligible compared to the activity in the market, while values of $\alpha$ close to one mean that most of the activity in the market came from the transactions of the hidden order. }
\end{itemize}

Summarizing, we expect that the market impact of a hidden order is a function 
\begin{equation}
r = f(N,V,T,f_{mo}),
\end{equation}
plus possibly other variables specific of the stock, such as the participation rate, the capitalization, the volatility, or the spread.
We will now try to simplify this by understanding some of the relationships between the dependent variables and by conditioning on some of them in our analysis.  Note that in all the analyses and figures we compute error bars as standard errors.  It should be born in mind that this procedure underestimates the errors due to the heavy tails of the fluctuations and due to possible long-memory properties of the data. 

\section{Statistical properties of hidden orders}\label{sec:statistics}

\begin{figure}
  \includegraphics[width=8.5cm,angle=0]{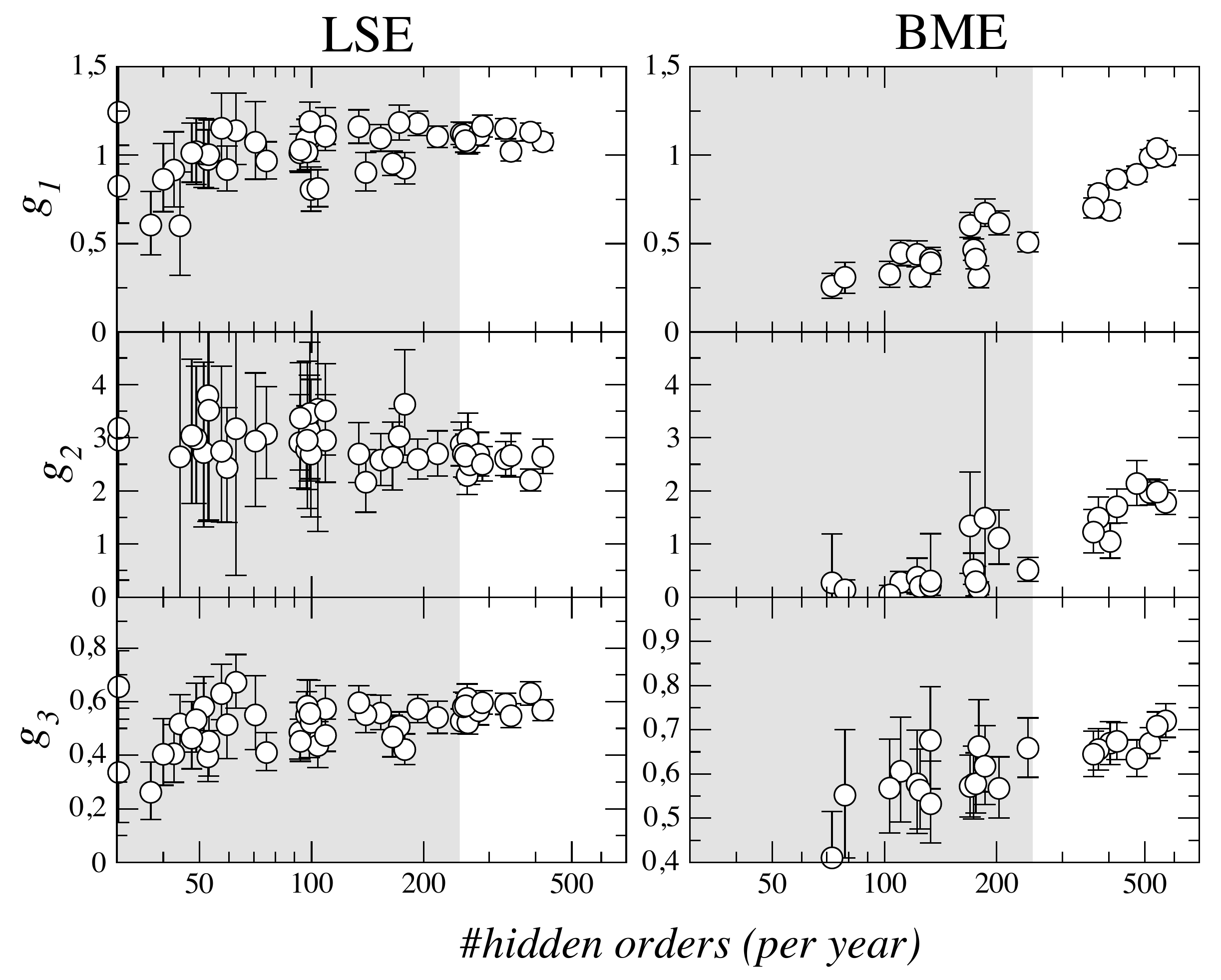}
\caption{Exponents $g_i$ ($i=1,2,3$) of the allometric relations of Eq. \ref{allometric} for each of the stocks considered in our LSE and BME databases and for hidden orders with $N\ge 10$ and $T< 1$ day, as a function of the number of detected hidden orders per year. Error bars are 95\% confidence intervals obtained by bootstrapping the data. In the analysis of market impact we consider only stocks with at least $250$ hidden orders per year (those in the white area of the figure).}
\label{g1g2g3}
\end{figure}

We investigate the statistical properties of the variables characterizing hidden orders. Ref. \cite{vaglica} considered a set of $3$ most capitalized stocks traded at the BME and studied the probability distribution of the variables characterizing the hidden orders and the scaling relations between these variables. In Ref. \cite{vaglica} no restriction on the length or on the fraction of market orders was set on the hidden orders. The authors of Ref. \cite{vaglica} found that the distribution of hidden order size is fat tailed and consistent with a distribution with infinite variance. They also showed that this broad distribution is due to an heterogeneity of scales among different brokerage firms rather than to the heterogeneity of scales within the hidden orders of each brokerage firm. By using Principal Component Analysis (PCA) on the logarithm of the variables characterizing the hidden orders, it was found that $N$, $V$ and $T$ are related through scaling relationships
\begin{equation}
N \sim V^{g_1},\quad T \sim V^{g_2},\quad N \sim T^{g_3}.
\label{allometric}
\end{equation}
where $g_1 \simeq 1$, $g_2 \simeq 2$ and $g_3 \simeq 0.66$ for 3 highly capitalized stocks in the BME and including all hidden orders. We repeat the two dimensional PCA analysis of \cite{vaglica} on our much larger data set. 
Figure \ref{g1g2g3} shows the value of the three exponents for all the stocks as a function of the number of hidden orders per year. We observe that for stocks with a small number of hidden orders the heterogeneity in the value of the exponents is pretty large, while, as the number of hidden orders detected by the algorithm increases, the exponent estimations become less noisy and tend to converge to similar values. Moreover for BME stocks there is a clear trend of the exponents as a function of the number of hidden orders.
In order to measure market impact in a statistically reliable way, we pool together data from different stocks. We need therefore an homogeneous sample of stocks. To this end in the following analysis we restrict our dataset to those stocks for which our algorithm detects at least 250 orders per year. These stocks are TEF, SAN, BBVA (as in \cite{vaglica}) but also REP, ELE, IBE, POP and ALT for the BME market and AZN, BSY, CCH, DVR, GUS, KEL, PO, PSON, SIG, TATE and TSCO for the LSE market. Moreover, in this paper we will focus mainly on short hidden orders, considering the set of hidden orders of time duration $T$ smaller than one trading day. The reason for this choice, detailed below, is to obtain stable statistical averages for the market impact. Applying these two restrictions, we obtain a final dataset that contains 14,655 hidden orders in the BME and 11,165 orders for the LSE (see Table \ref{table1}). 
We repeat the two-dimensional PCA analysis of \cite{vaglica} on the pooled set of hidden orders from different stocks. We find for the BME market the following exponents
\begin{eqnarray}
g_1 & =& 0.81\ \ (0.79;0.82),\\
g_2 & = &1.57\ \ (1.43;1.72),\\
g_3 &= &0.67\ \ (0.65;0.68),
\end{eqnarray}
where quantities in parenthesis are 95\% confidence intervals obtained through bootstrapping the data. These relations explains 83\%, 61\% and 80\%, respectively, of the variance observed in the data. For the  LSE dataset we get
\begin{eqnarray}
g_1 & =& 0.99\ \ (0.98;1.01), \\
g_2 &=& 2.41\ \ (2.29;2.52), \\
g_3 &=& 0.58\ \ (0.57;0.59),
\end{eqnarray}
and these relations explain 88\%, 75\% and 86\%, respectively, of the variance. These allometric relations are roughly consistent with those obtained in Ref. \cite{vaglica}.

\begin{table}[t]
\caption{Statistics of the hidden order ensembles used in the paper. Only hidden orders with $T < 1$ day and $N > 10$ transactions are used.}
\begin{ruledtabular}
\begin{tabular}{l|cccccc}
Market & \# orders & $\langle N \rangle$ & $\langle f_{mo} \rangle$ & $\langle \alpha \rangle$ & $\langle R \rangle$ & $\langle R \rangle_{f_{mo}>0.8}$ \\ \hline
BME & 14,655 & 95.58 & 0.52 & 0.17  & 1.127 & 3.983 \\
LSE & 11,165 & 97.53 & 0.53 & 0.34  & 0.587 & 2.156
\label{table1}
\end{tabular}
\end{ruledtabular}
\end{table}

\begin{figure}
  \includegraphics[width=8.5cm,angle=0]{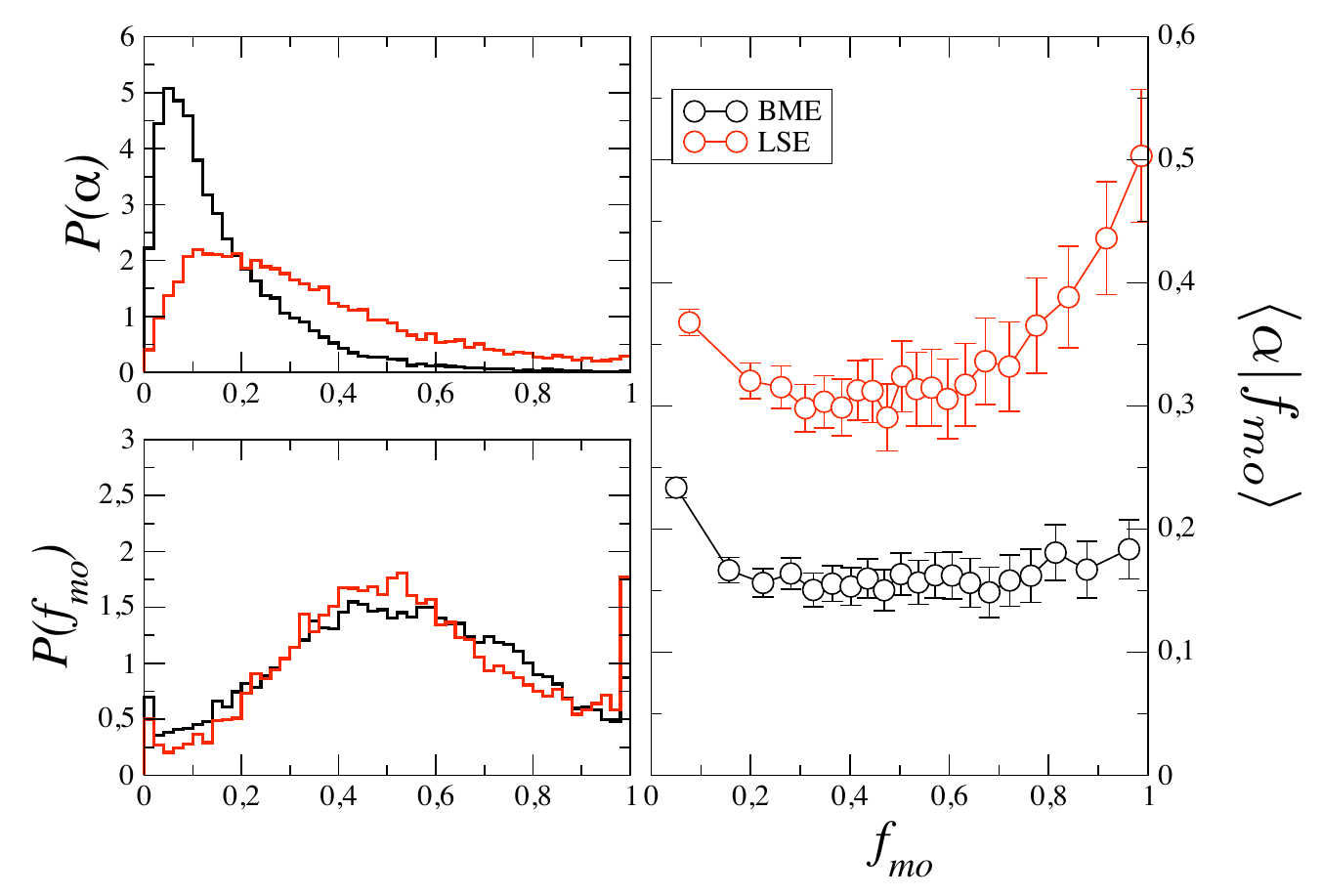}
\caption{Ensemble statistics of the fraction of market orders $f_{mo}$ and participation rate $\alpha$ of hidden orders in both the BME and LSE. Left panels show the probability distribution function of both parameters, while the right panel shows the conditional average of the participation rate conditioned on a given value of $f_{mo}$.}
\label {alphafmo}
\end{figure}

 The left panels of Fig. \ref{alphafmo} show the probability density function of $f_{mo}$ and of the participation rate $\alpha$. We observe that the distribution of the fraction of market orders is rather broad and is roughly centered around $f_{mo}=0.5$.  In addition two peaks are observed for $f_{mo}\simeq0$ and $f_{mo}\simeq1$. For the BME the participation rate has a peak around $\alpha=5\%$, while for the LSE the distribution of $\alpha$ is broader, and peaks at a value closer to $20\%$.  This value is pretty large and we do not have an explanation for the difference in the participation rate between the two markets. Finally, the two parameters $\alpha$ and $f_{mo}$ are not independent. Figure \ref{alphafmo} shows the expected value of $\alpha$ conditioned on $f_{mo}$. For the BME the expected value of $\alpha$ is almost constant except for very small values of $f_{mo}$.  In contrast, for the LSE the dependence is much stronger.  The participation rate is higher when $f_{mo}$ is at either of its extremes.  

The bottom left panel of Fig. \ref{alphafmo} shows that $f_{mo}$ has a broad distribution. Hidden orders can therefore differ a lot in terms of the fraction of market orders used to complete them. In the investigation of the market impact of hidden orders we will consider hidden orders characterized by a restricted set of values of $f_{mo}$ to better characterize their profile with respect to the fraction of market orders used to complete the hidden order. Specifically we will use $f_{mo} > 0.8$ (large fraction of market orders used) and  $f_{mo} < 0.2$ (large fraction of limit orders used).  The reasons why we expect this distinction to be critically important will be described in the next section.



\section{Market impact}\label{sec:impact}

\subsection{Definition}
The main focus of this paper is the empirical measurement of the market impact of hidden orders. 
Given a hidden order traded on stock $i$ between times $t$ and $t+T$, we measure the market impact by considering the change in the log price of the stock  between time $t$ and time $t+T$, i.e.
\begin{equation}
 r_i(t,T) = \log p_{i,t+T}-\log p_{i,t}, 
 \end{equation}
 where $p_{i,t}$ is the price of stock $i$ at time $t$. We have used for $p_i$ the midprice, but our results do not depend on this. Our objective is to study how $r_i(t,T)$ changes as a function of the main properties of the hidden order.
Different stocks have different scales of their price fluctuations. In order to be able to take the average of market impact across different stocks, we rescale it by dividing by the mean value of the spread $s_i$ of the stock during the year, where the spread is the difference between the lowest selling price (ask) and the highest buying price (bid).  Specifically, we define the rescaled market impact as
\begin{equation}\label{signedreturn}
R_i(t,T) = \epsilon_i r_i(t,T)/s_i.
\end{equation}
where, as before, $\epsilon_i=+1$ for a buy hidden order and $\epsilon_i=-1$ for a sell hidden order.
Although we observe a small asymmetry between the market impact of buy vs. sell orders, similar to that observed elsewhere \cite{aitken}, for the purpose of our study here we lump together buy and sell hidden orders in order to obtain better statistics.

\subsection{The noisy nature of market impact}
 
 \begin{figure}
  \includegraphics[width=8.5cm,angle=0]{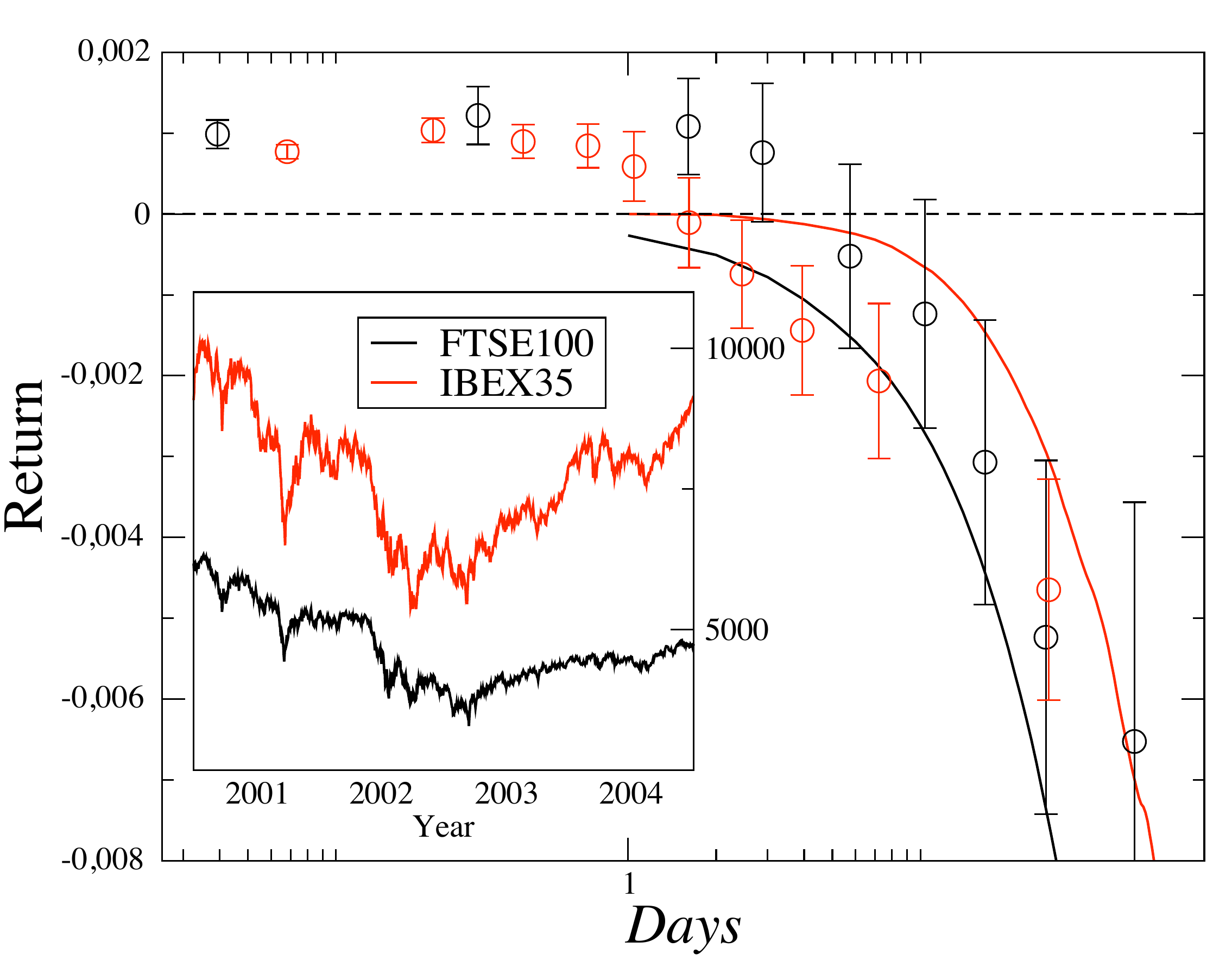}
\caption{Conditional average $\langle R | T \rangle$ of the rescaled impact of hidden orders (Eq.\ (\ref{signedreturn})) as a function of their time duration $T$ (symbols) compared to the average return of the stock market index over random periods of the same time duration (solid lines). The inset shows the price of the FTSE100 and IBEX35 indices over the period of study. In this figure we are using all detected hidden orders without any conditioning on $T$ or $f_{mo}$ values but with $N>10$.}
\label{fort}
\end{figure}

While a given hidden order is trading there are typically many other orders trading at the same time, as well as news arrival, and thus there is a considerable amount of noise in the price change associated with any particular hidden order.  The price change associated with a hidden order functionally depends on several factors, which can be written
\begin{equation}\label{model}
 r_i(t,T) = {\cal R}[r_{M}(t,T),\rho_i(t,T),\eta_i(t,T)].
\end{equation}
where $r_M$ corresponds to market-wide movements \cite{lakonishok}, $\rho$ is the average market impact of the hidden order, and $\eta_i$ is the background uncorrelated noise coming from the trading of the rest of the market \cite{plerou}. 
While the background noise can be controlled by taking averages over different orders with the same properties or restricting our analysis to very small values of $T$, market-wide movements remain large, especially for large values of $T$. During the years 2001-2004 stock markets were in a substantial decline for more than two years, only recovering at the end of 2003 and 2004 (see the inset of Figure \ref{fort}). 

Figure \ref{fort} shows the conditional average $\langle R | T \rangle$ of the rescaled market impact of the hidden orders as a function of their time duration $T$. We observe that for $T$ larger than one day, rescaled impact is on average negative,  irrespectively of the sign of the hidden order. The reason for this phenomenon is that market-wide movements were mostly negative for values of $T$ larger than one day. Only for hidden orders of duration close to or below one day do we observe negligible changes in market indexes when compared to price changes during hidden order completion. This is the main motivation of our choice of restricting our study to hidden orders of duration $T$ less or equal to one day\footnote{Other authors \cite{lakonishok} have proposed to use industrial sector indexes as proxies for market-wide movements of a given stock and thus the study can be extended to larger values of $T$. We do not follow this procedure due to lack of that information.}. 

\subsection{Impact of limit orders vs. market orders}

It is important to stress that market impact comes about through changes in supply and demand, and that this causes a strong {\it a priori} difference in the impact one expects to observe in the execution of a limit order vs. a market order.  For example consider buy orders.  A buy market order reflects an increase in demand at the current price.  If sufficiently large it will cause a positive price change.  Since in a continuous double auction  market orders always execute against limit orders, this implies that the sell limit order that the buy market order executes against will generate a positive market impact.  We therefore expect that executed limit orders have the opposite impact of market orders:  Buying drives the price down and selling drives it up.

The problem with this line of reasoning is that we are considering only {\it executed} limit orders, which creates a strong selection bias.  To measure the impact of limit orders correctly we need to condition on all orders that are placed, rather than only on those that are executed.  When this is done the impacts for limit orders should be roughly the same as for market orders, as otherwise it would be possible to make a profit by simply using limit orders instead of market orders.  If a buy limit order is placed below the current price it is executed only if the price drops.  The probability of execution of a limit order depends on future price movements:  under an adverse price movement the probability of execution is higher than for a favorable price movement.  This is caused in part by the mechanical dynamics of a random walk, but also by asymmetric information:  Placing a limit order gives others the option of executing at their will, when they have information that indicates it is favorable to do so.  This phenomenon is called {\it adverse information}.  When this is properly taken into account, limit orders have impact in the direction one would expect, i.e. buying has positive impact and selling has negative impact \cite{Zamani09,Eisler09}.   Furthermore the magnitude of the impact of limit orders when the selection effects are properly taken into account is comparable to that of market orders.

For the BME we have a record of transactions but not of orders.  Thus to measure market impact and avoid the selection bias associated with executed limit orders we are forced to use only those hidden orders that are predominantly built out of market orders.  For consistency we analyze both the BME and the LSE data in the same way.

In Table \ref{table1} we show the mean value of the rescaled market impact $R$ of Eq. (\ref{signedreturn}) for hidden orders of duration less than one day. We also show the mean value $\langle R \rangle_{f_{mo}>0.8}$ of the rescaled market impact computed over the set of hidden orders with a large fraction of market orders ($f_{mo}>0.8$). $\langle R \rangle_{f_{mo}>0.8}$ is significantly larger than $\langle R \rangle $ indicating that hidden orders mainly composed by market orders have on average a larger market impact than hidden orders composed of both limit and market orders.  


\subsection{Impact vs. $N$}

\begin{figure}
      \includegraphics[width=8.5cm,angle=0]{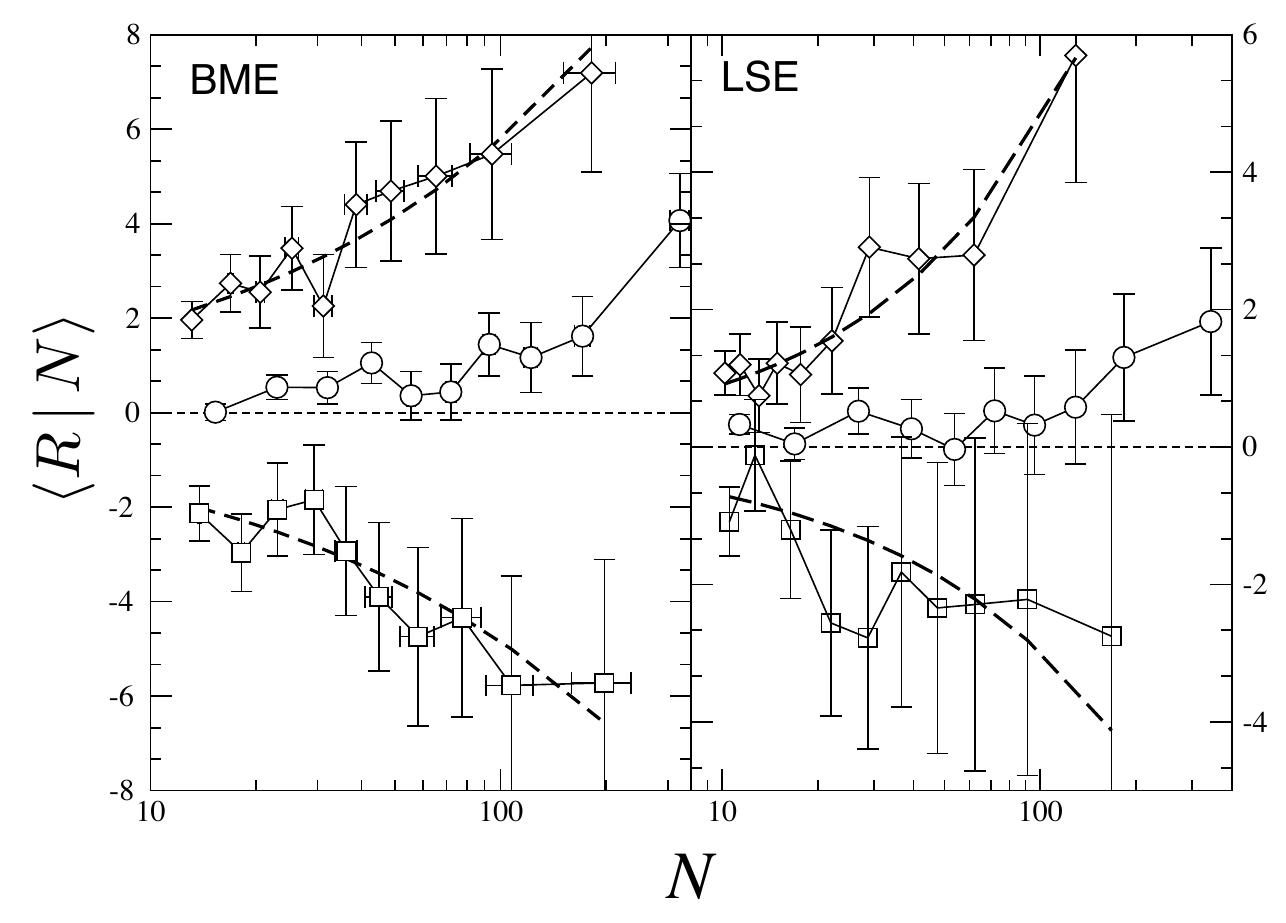}
    \caption{
Average rescaled market impact $R$ for hidden orders shorter than 1 day as a function of $N$ for the BME (left) and LSE (right). Circles are the results for all hidden orders, while squares are the results when there is a low fraction of market orders ($f_{mo} < 0.2$) and diamonds are for when there is a large fraction of market orders ($f_{mo} > 0.8$). Dashed lines are power law fits $R \sim N^\gamma$. Values of $\gamma$ are reported in Table \ref{table2}. 
  }
  \label{rnfig}
\end{figure}

Figure \ref{rnfig} shows the average over all hidden orders of the rescaled market impact $\langle R | N\rangle$ as a function of the conditioning variable $N$.  This grows slightly as a function of $N$, but one must keep in mind that the meaning of this is difficult to interpret in view of the discussion above, since we are averaging together a roughly equal number of market orders and executed limit orders.

To investigate the average market impact and minimize the effect of the selection bias, we divide the data into two groups: liquidity providing hidden orders, with $f_{mo} < 0.2$, and liquidity demanding hidden orders, with $f_{mo} > 0.8$.  As expected, for the former group the market impact is on average negative, while for the latter it is positive.  Using ordinary least squares, we find that for both groups the dependence of  $\langle R | N\rangle$ on $N$ is well described by the power law
\begin{equation}\label{powerlaw}
|\langle R|N\rangle| = A\ N^\gamma.
\end{equation}
The estimated parameters are in Table \ref{table2}.
In summary, we find that the market impact of hidden orders dominated by market orders is consistent with 
\begin{equation}\label{eq:impact}
\langle r | N \rangle \propto \epsilon s N^\gamma
\end{equation}
where $\epsilon$ is the sign of the order and $s$ is the spread. For hidden orders dominated by limit orders the market impact is very similar to minus the impact of hidden orders dominated by market orders. 

For the BME the exponent is consistent with a square root function while for the LSE the exponent of the impact is slightly larger than $0.5$. The square root dependence is consistent with other studies and with the predictions of some models. The BARRA model \cite{barra} uses an exponent $0.5$ for estimating market impact. Almgren et al. \cite{almgrenrisk} found an exponent approximately equal to $0.6$ for the temporary impact of hidden orders. The theories of references \cite{gabaix} and \cite{henri} predict that the exponent of the impact should be roughly $0.5$, with the exact value depending on the heavy tail of the volume distribution.

\begin{table}[t]
\caption{Parameters of the fitting of the market impact with Eq. \ref{powerlaw}.} 
\begin{ruledtabular}
\begin{tabular}{l|ll|ll}
Market & $A_{f_{mo}>0.8}$ & $\gamma_{f_{mo}>0.8}$& $A_{f_{mo}<0.2}$ & $\gamma_{f_{mo}<0.2}$\\ \hline
BME & $0.63\pm0.17$& $0.48\pm0.07$& $-0.63\pm0.22$&$0.44\pm0.09$\\
LSE & $0.17\pm0.05$&$0.72\pm0.10$&$-0.16\pm0.14$&$0.64\pm0.30$
\label{table2}
\end{tabular}
\end{ruledtabular}
\end{table}

In Eq. (\ref{eq:impact}) the spread gives the proportionality constant, i.e. the global scale of the impact. By using the results of Wyart et al. \cite{wyarth}, who derive a proportionality between the spread and the volatility per trade, it is possible to rewrite Eq. \ref{eq:impact}  in such  a way that the proportionality constant is the volatility per trade. 

\subsection{Temporary vs. permanent impact}

\begin{figure} [t]
      \includegraphics[width=8.5cm,angle=0]{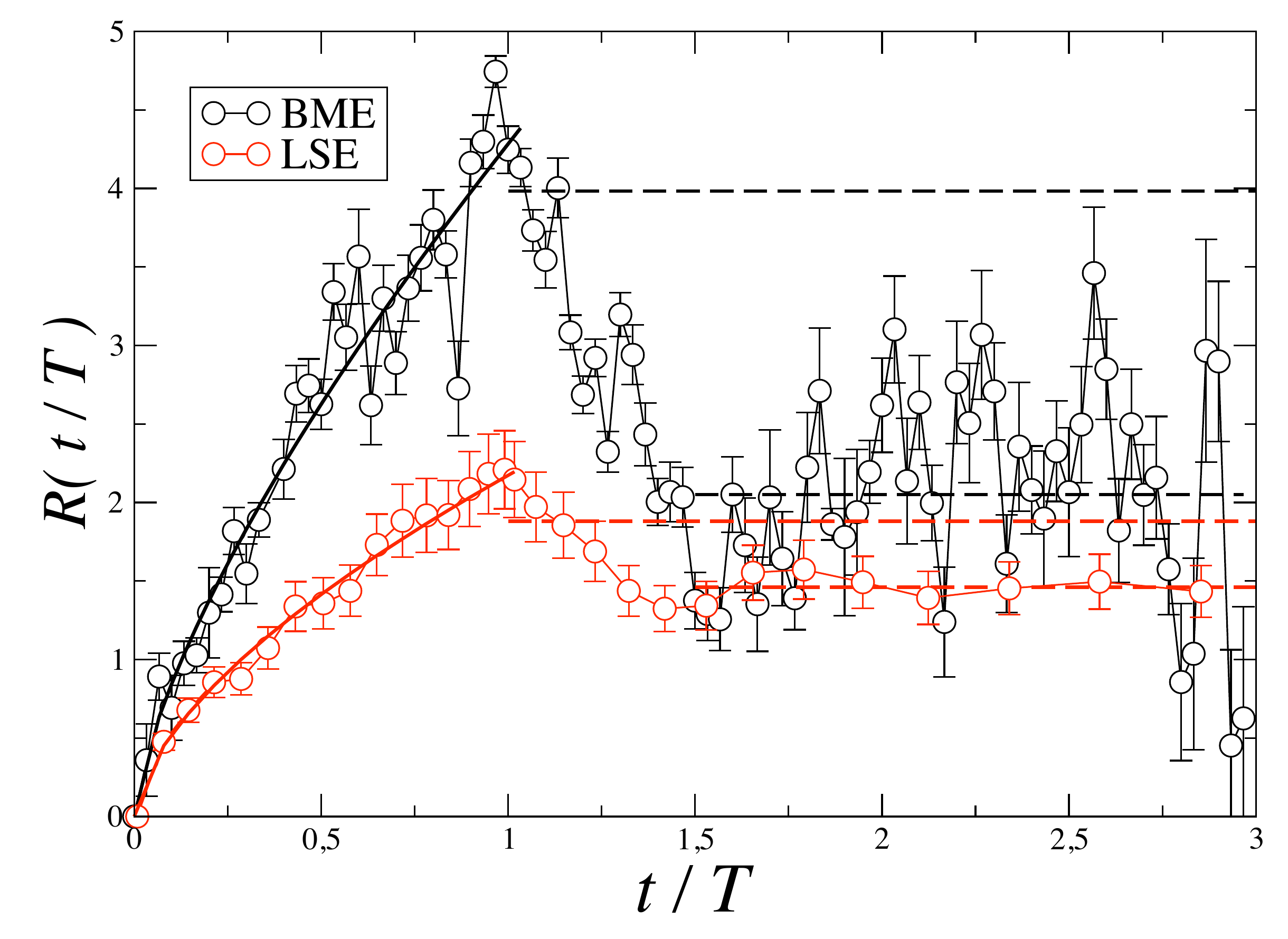}
    \caption{ Market impact versus time.  The symbols are the average value of the market impact of the hidden order as a function of the normalized time to completion $t/T$.  The rescaled time $t/T = 0$ corresponds to the starting point of the hidden order, while $t/T$ = 1 corresponds to the end of the hidden order. The rescaled time $t/T$ is extended up to $t/T = 3$ to study the permanent and temporal impact of the hidden order. Solid lines are power-law fits (see text) while dashed lines correspond to temporary (upper) and permanent (lower) market impact. Temporary impact $R_{temp}$ is measured at the end of the order $t = T$ (see Table \ref{table1}), while permanent impact $R_{perm}$ is obtained through an average of the $R(t/T)$ with $1.5 \leq t/T \leq 3$ obtaining $R_{perm} = 2.03\pm 0.68$ for the BME and $R_{perm} = 1.48\pm 0.06$ for the LSE. Data is only for hidden orders with $f_{mo} > 0.8$. 
  }
  \label{impact}
\end{figure}

Finally we study how market impact builds as hidden orders are executed and how the price reverts when the execution is completed. Here we show the impact as a function of time rather than of executed trades. In order to consider hidden orders of different length we normalize the time by dividing it by the execution time $T$. With our normalized time, the initial time of the order corresponds to $t/T =0$ while the final time is $t/T = 1$.  We consider only orders with $f_{mo}>0.8$.  The results are shown in Figure \ref{impact} where we see that earlier transactions within the hidden order have more impact than later ones. In fact we observe that 
\begin{eqnarray}
&R\sim (4.28 \pm 0.21) \times  \left(\frac{t}{T}\right)^{0.71\pm 0.03}~~~~~~~(BME)\\
&R\sim (2.13\pm 0.05) \times  \left(\frac{t}{T}\right)^{0.62\pm 0.02} ~~~~~~~(LSE)
\end{eqnarray}


We also observe that after the completion of the hidden order price drops, suggesting that not all of the market impact is permanent. The tendency for reversion has also been observed previously \cite{keim,gerig, lakonishok}.  We compare the impact at its peak when the order has just finished, $R_{temp} = R(t = T)$ (see Table 1), to the permanent impact $R_{perm} = R(t \gg T)$ estimated by averaging over a period $1.5 \leq t/T \leq 3$.   The drop in impact is $R_{perm}/R_{temp} \simeq 0.51\pm 0.22$ (BME) and $R_{perm}/R_{temp} \simeq 0.73 \pm 0.18$ (LSE).

It is interesting to note that the square root impact and the reversion can be predicted through a simple argument based on the hypothesis that the price after reversion is equal to the average price paid during execution \cite{henri}. If during execution price impact grows like $ A \times (t/T)^\beta$ then the average price payed by the agent who executes the order is
\begin{equation}\label{integral}
\langle p \rangle= p_t+ A \int_0^1 (t/T)^\beta dt= p_t+\frac{A}{1+\beta},
\end{equation}
i.e. the permanent impact is $1/(\beta+1)$ of the peak impact. In our case and using the exponents $\beta$ obtained in figure \ref{impact} we get $1/(\beta +1) \simeq 0.58\pm 0.01$ for the BME and $1/(\beta + 1) \simeq 0.62\pm 0.02$ for the LSE which are statistically similar to the ratios $R_{perm}/R_{temp}$ for each market.

\section{Trading profile}\label{sec:profile}
In this section we investigate how hidden orders are executed as a function of time, which we call the trading profile.  By this we mean the traded volume of the hidden order as a function of time elapsed from the time of the first trade. As before, in order to average across orders of different length we use the normalized time $t/T$. We measure the normalized average volume of each transaction $v_i / \langle v_i \rangle$ traded at time $t$ inside the hidden order as a function of the normalized time.  Here $\langle v_i \rangle$ is the average volume exchanged in the individual transactions used to execute the hidden order. 

\begin{figure} [t]
      \includegraphics[width=8.5cm,angle=0]{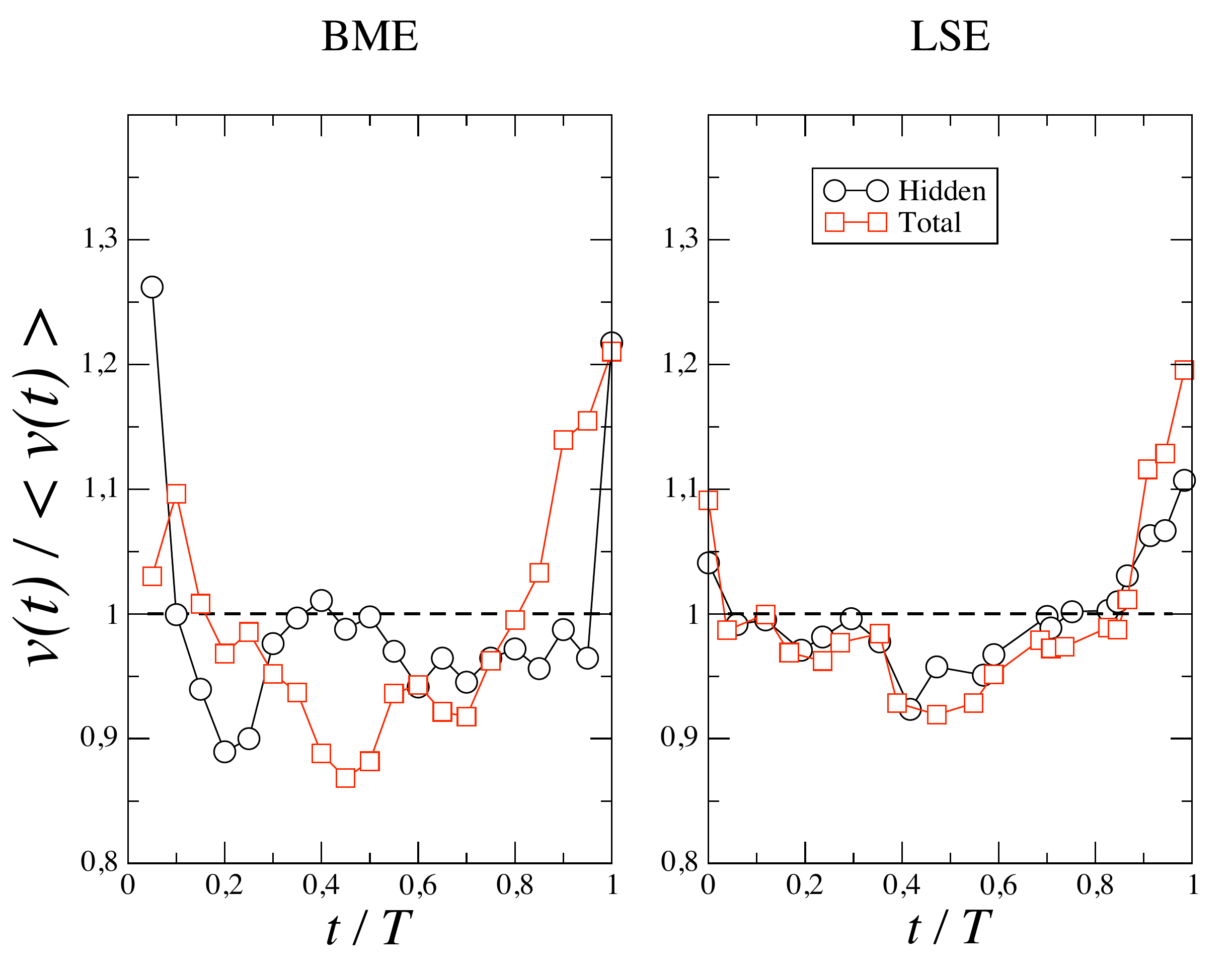}
    \caption{ Trading profile inside the hidden order. Average volume of the transactions within the hidden order divided by the average volume in the hidden order as a function of the normalized time $t/T$. Circles are the results for all hidden orders, while squares are the volume traded in the market (in the same stock) concurrently with the hidden order. Data is only for hidden orders with $f_{mo} > 0.8$.
 }
  \label{time}
\end{figure}

Figure \ref{time} shows that trading within the hidden order is fairly homogeneous except for the initial and final times of the order, for which there is a small increase in the traded volume. This can be understood if we look at the concurrent trading in the market. In Figure \ref{time} we see that the profile of the hidden order substantially matches the concurrent trading in the market.  In fact, the rise and fall of concurrent trading is a bit stronger than it is for hidden orders.  The cause and effect of this phenomenon is not clear:  Does trading rise and fall because of the pattern of hidden order placement, or do people placing hidden orders try to match trading volume, e.g. through VWAP (volume weighted average price) strategies?

\begin{figure}[t]
      \includegraphics[width=8.5cm,angle=0]{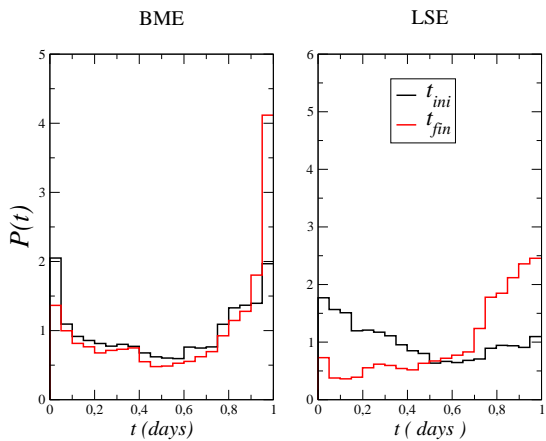}
    \caption{ Initial and final times of  the hidden orders. Probability distributions of the initial time $t_i$ and final time $t_f$ of the hidden orders, measured with respect of the time of the day.  Data is only for hidden orders with $f_{mo} > 0.8$
  }
  \label{insets}
\end{figure}

As shown in Figure \ref{insets}, the starting and ending of hidden orders is substantially correlated with overall volume of trading in the market.  In particular we see that 
a significant fraction of hidden orders start at the beginning of the day and finish at the end of the day. These are the times of day in which the volume traded is larger, corresponding to the well-known U-shaped volatility and trading volume profile.
Once again the cause and effect is not clear. It has been shown that when impact of a transaction decays gradually with time, the optimal trading strategy has a turnpike shape \cite{hasbrouckbook,obizhaeva,bfl}.  Thus this might drive the end of day increase in trading, or alternatively, the daily variation in trading may drive the profile of hidden orders simply due to the desire to match traded volume.

\section{Conclusions}\label{sec:conclusions}

In this paper we have empirically studied the main properties of the impact and of the trading protocol of intradaily hidden orders using a large fraction of either market orders or limit orders.
We have found that the temporary impact of hidden orders is concave and roughly described by a square root function of the hidden order size. Moreover the price reverts after the completion of the hidden order in such a way that the permanent impact is equal to roughly $0.5-0.7$ of the temporary impact.

We have also studied how the order is completed in time and we have shown that more volume of the hidden order is traded at the beginning and at the end of the hidden order.  When we take into account that hidden orders are more likely to start at the beginning of a day and are more likely to end near the end of the day, this roughly matches the volume traded in the market.

The fact that we observe similar behavior in both the London and Spanish stock exchanges, and that others have also observed this in the New York Stock Exchange, suggests the possibility of a ``law" for market impact.  It will be very interesting to see whether this hypothesized law continues to hold up under future studies.


\acknowledgments 
FL and JDF acknowledge Bence Toth and Henri Waelbroeck for useful discussion.
GV, FL, and RNM acknowledge financial support from the PRIN project 2007TKLTSR  ``Computational markets design and agent-based models of trading behavior".
EM, GV, FL, and RNM acknowledge Sociedad de Bolsas for providing the data and the Integrated Action Italy-Spain ``Mesoscopics of a stock market" for financial support.  EM acknowledges partial support from MEC (Spain) through grants Ingenio-MATHEMATICA and MOSAICO and
Comunidad de Madrid through grant SIMUMAT-CM. JDF, JV and AG would like to thank Barclays Bank, Bill Miller, and National Science Foundation grant 0624351 for support.  Any opinions, findings, and conclusions or recommendations expressed in this material are those of the authors and do not necessarily reflect the views of the National Science Foundation.

\end{document}